\documentclass[11pt, onecolumn, journal]{IEEEtran}
\usepackage{amssymb}
\usepackage{amstext}
\usepackage{latexsym}
\usepackage{times}
\usepackage{amssymb}
\usepackage{amsmath}
\usepackage{epsfig}
\usepackage{cite}
\usepackage{verbatim}
\usepackage{enumerate}
\usepackage{subfigure}
\usepackage{multicol}
\usepackage{pstricks}
\usepackage{pst-node}
\usepackage{setspace}
\usepackage{graphicx}
\usepackage{picinpar}
\newtheorem{theorem}{Theorem}%[section]

\newtheorem{proposition}{Proposition}%[section]

\renewcommand{\baselinestretch}{1.85}

\begin{document}
\title{Optimal Relay Functionality for SNR Maximization in Memoryless Relay Networks}
\author{\authorblockN{Krishna Srikanth Gomadam, Syed Ali Jafar}\\
\authorblockA{Electrical Engineering and Computer Science\\
University of California, Irvine, CA 92697-2625\\
Email: {kgomadam@uci.edu, syed@ece.uci.edu}}} \maketitle
\IEEEpeerreviewmaketitle

\begin{abstract}

We explore the SNR-optimal relay functionality in a
\emph{memoryless} relay network, i.e. a network where, during each
channel use, the signal transmitted by a relay depends only on the
last received symbol at that relay. We develop a generalized notion
of SNR for the class of memoryless relay functions. The solution to
the generalized SNR optimization problem leads to the novel concept
of minimum mean square uncorrelated error estimation(MMSUEE). For
the elemental case of a single relay, we show that MMSUEE is the
SNR-optimal memoryless relay function regardless of the source and
relay transmit power, and the modulation scheme. This scheme, that
we call estimate and forward (EF), is also shown to be SNR-optimal
with PSK modulation in a parallel relay network. We demonstrate that
EF performs better than the best of amplify and forward (AF) and
demodulate and forward (DF), in both parallel and serial relay
networks. We also determine that AF is near-optimal at low transmit
power in a parallel network, while DF is near-optimal at high
transmit power in a serial network. For hybrid networks that contain
both serial and parallel elements, and when robust performance is
desired, the advantage of EF over the best of AF and DF is found to
be significant. Error probabilities are provided to substantiate the
performance gain obtained through SNR optimality. We also show that,
for \emph{Gaussian} inputs, AF, DF and EF become identical.
\end{abstract}

\newpage
\section{Introduction}
The traditional wireless communication problem is to design
effective coding and decoding techniques to enable reliable
communication at data rates approaching the capacity of a channel.
The channel is defined by a set of \emph{given} assumptions
regarding the physical signal propagation environment between the
transmitter and the receiver. However, recent focus on cooperative
communications presents a remarkable change of paradigm where in
addition to the physical environment, \emph{ the network is the
channel} \cite{break_net}. In other words, with cooperative
communications the effective channel between the original source and
the final destination of a message depends not only on the given
physical signal propagation conditions but also the signal
processing at the cooperating nodes. The change of paradigm is quite
significant. With cooperative communications, not only is there a
need to optimally design the encoder and decoder at the source and
destination, but also to \emph{design the channel} itself by
optimally choosing the functionality of the intermediate relay
nodes. The choice of relay function is especially important as it
directly affects the potential capacity benefits of cooperation
which have been shown to be quite significant \cite{sendonaris1,
sendonaris2, cap_theorems_relay_kramer,
cap_wireless_relay_hostmadsen, cap_cheap_relay}.

A number of relay strategies have been studied in literature. These
strategies include amplify-and-forward \cite{laneman_worell_exploit}
\cite{laneman_mod_det}, where the relay sends a scaled version of
its received signal to the destination, demodulate-and-forward
\cite{laneman_mod_det} in which the relay demodulates individual
symbols and retransmits, decode-and-forward
\cite{laneman_efficientprotocols} in which the relay decodes the
entire message, re-encodes it and re-transmits it to the
destination, and compress-and-forward \cite{cap_theorems_cover}
\cite{cap_theorems_relay_kramer} where the relay sends a quantized
version of its received signal. In
\cite{distributed_mmse_nima_sayed}, gains are determined for AF
relays to minimize the MMSE of the source signal at the destination.
It is shown that significant savings in power is achieved if there
is no power constraint on the relays. Similarly in
\cite{distributed_multiuser_mmse_berger_wittenben} gains for AF
relays in a multiuser parallel network are determined that realizes
a joint minimization of the MMSE of all the source signals at the
destination.

From a practical standpoint, the benefits of cooperation are offset
by the cost of cooperation in terms of the required processing
complexity and transmit power at the relay nodes. The complexity of
the signal processing at the relay could range from highly
sophisticated decode-and-forward or compress-and-forward techniques
\cite{three_node_elgamaal} that require joint processing of a long
sequence of received symbols to memoryless schemes such as
amplify-and-forward or demodulate-and-forward that process only one
symbol at a time. Clearly, the most desirable schemes are those that
approach the limits of cooperative capacity with minimal processing
complexity at the relays. Memoryless relay functions are highly
relevant for this objective. In addition to their simplicity,
memoryless relays are quite powerful in their capacity benefits. For
example, the memoryless scheme of amplify-and-forward is known to be
the capacity-optimal relay scheme for many interesting cases \cite{
cap_lar_gauss_relay, gastpar_infocom, elgamaal_linear, break_net,
relay_without_delay}. The effect of finite block-length processing
at the relay on the capacity of serial networks is analyzed in
\cite{line_networks, line_nets2}. In \cite{cap_icc06}, the
memoryless MMSE estimate and forward scheme has been shown to be
capacity optimal for a single relay system. For a single relay and
with BPSK modulation, the BER-optimal memoryless scheme is found by
Faycal and Medard \cite{medard_optimal_fn}. The BER-optimal relay
function turns out to be a Lambert W function normalized by the
signal and noise power.
%It is shown in \cite{line_networks, line_networks2} that most of the
%capacity of a linear network is achieved with finite processing at
%the relays.

In this paper we explore the SNR-optimal signal processing function
for memoryless networks with possibly multiple relays. While SNR
optimality does not always guarantee capacity or BER optimality, it
is a practically useful performance metric. SNR-optimization is
especially interesting for its greater tractability that allows
analytical results where capacity and BER optimizations may be
intractable, e.g. with multiple relays.

\subsection{Notations}
Throughout the paper, $\mathcal{E}[.]$ denotes the standard
expectation operator, $^*$ represents the conjugation operation.
$|.|$ and $\mathfrak{Re}(.)$ denotes the absolute and real part of
the argument respectively.
\section{Shaping the Relay Channel: Amplification, Demodulation and Estimation}
\label{shape} In this section, we discuss the relay functions of
common memoryless forwarding strategies and provide new perspectives
that lead us to a novel and superior memoryless forwarding
technique.
\subsection{Soft and Hard Information: Amplify and Demodulate}
Within the class of memoryless relay strategies, amplification and
demodulation are the most basic forwarding techniques
\cite{laneman_mod_det}. An AF relay simply forwards the received
signal $r$ after scaling it down to satisfy its power constraint.
The relay function for AF can be written as
\begin{equation}
f_{AF}(r)=\sqrt{\frac{P_{R}}{P+1}}r.
\end{equation}
Evidently with AF, the relay tries to provide soft information to
the destination. A disadvantage with this technique is that
significant power is expended at the relay when $|r|$ is high. In DF
schemes, demodulation of the received symbol at the relay is
followed by modulation with its own power constraint $P_{R}$. For
BPSK modulation, the relay function for DF can be expressed as
\begin{equation}f_{DF}(r)=\sqrt{P_{R}}\mbox{sign}(r),\end{equation} where
$\mbox{sign}(r)$ outputs the sign of $r$. Due to demodulation, the
relay transmitted signal carries no information about the degree of
uncertainty in the relay's choice of the optimal demodulated symbol.
Demodulation at the relays can lead to severe performance
degradation in some scenarios. For example, in a parallel relay
network, reliability information can be utilized to achieve better
performance over DF.

From the relay functions of AF and DF, one can argue that an optimal
relay function should provide soft information when there is an
uncertainty in the received symbol, and at the same time should not
expend a lot of power when the cost of power out-weighs the value of
soft information.
\subsection{Estimate and Forward: A Novel Memoryless Forwarding Strategy}
The forwarding schemes can also be related to the fundamental signal
processing operations: \emph{detection} and \emph{estimation}. In
DF, the relay demodulates the received symbol employing MAP
detection rule, which is the optimal detection technique. So a DF
function can be viewed as a MAP detector followed by a modulator. In
a similar vein, AF can be viewed as a linear MMSE\footnote{Because
of the normalization associated with the relay power constraint all
linear estimates are equivalent} estimation scheme followed by
normalization to satisfy the relay power constraint.
\[f_{AF}(r)=\beta^{'}\widehat{X}_{linear}(r)\]
where the linear estimate $\widehat{X}_{linear}(r)$ obtained at the
relay is given by
\[\widehat{X}_{linear}(r)=\frac{P}{P+1}r,\] and
\[\beta^{'}=\sqrt{\frac{P_{R}(P+1)}{P^2}}.\]
Viewing AF as linear MMSE leads naturally to the forwarding scheme
of EF where the unconstrained MMSE estimate is forwarded. The
unconstrained MMSE estimator which minimizes the distortion is given
by
\[\widehat{X}(r)=\mathcal{E}(x|r).\]
\begin{figure}
\center \centerline{\psfig{figure=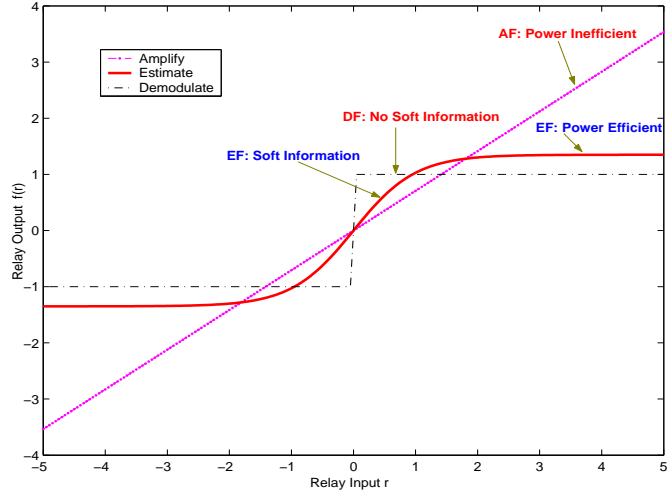,width=3.5in, height=2.6in}}
\caption{Relay function of common forwarding schemes (P=1)}
\label{confid}
\end{figure}
\subsubsection{Relay Function for EF with BPSK modulation}
When the source employs BPSK modulation, the MMSE estimate at the
relay is given by
\begin{equation}
\widehat{X}(r)=\sqrt{P}\tanh(\sqrt{P}r),
\end{equation}
where $\tanh(z)$ returns the hyperbolic tangent of $z$. The relay
function is therefore,
\[f_{EF}(r)=\sqrt{\frac{P_{R}}{\mathcal{E}[\tanh^2(\sqrt{P}r)]}}\tanh(\sqrt{P}r)\]

Note that all the memoryless schemes operate at sampled output of
the matched filter. In this regard all the schemes have similar
processing complexity. It is worth noting that while EF and AF
require amplitude digitization, DF does not. Figure \ref{confid}
shows the relay functions for AF, EF and DF for $P=1$. It can be
seen that the relay function $f_{EF}$ is linear for small values of
$|r|$. Its slope reduces gradually and ultimately becomes flat
similar to $f_{DF}$. The function $f_{EF}=\sqrt{P}\tanh(\sqrt{P}r)$
is intuitively appealing for the following characteristics.

\begin{itemize}
\item{Soft information in region of uncertainty.}\item{ Limited power in region of high power cost.}
\end{itemize}
The insights obtained in this section will be useful in
understanding optimum relay functionalities in a multiple relay
network. In the next section, we determine the optimal memoryless
strategy in a single relay network.

\section{Single Relay Channel}\label{PROB_DEF_SEC}
\subsection{Problem Statement}  Consider an
elemental relay channel model as shown in the figure below, in which
a single relay R assists the communication between the source S and
the destination D. Both S-R
and R-D links are assumed to be non-fading. \\
\begin{window}[0,c,{{$~~~~~~~~~~~~~~~~~~~$\begin{picture}(0,0)%
\includegraphics{relay.pstex}%
\end{picture}%
\setlength{\unitlength}{2565sp}%
\begingroup\makeatletter\ifx\SetFigFont\undefined%
\gdef\SetFigFont#1#2#3#4#5{%
  \reset@font\fontsize{#1}{#2pt}%
  \fontfamily{#3}\fontseries{#4}\fontshape{#5}%
  \selectfont}%
\fi\endgroup%
\begin{picture}(7040,618)(464,-2195)
\put(464,-2099){\makebox(0,0)[lb]{\smash{{\SetFigFont{9}{10.8}{\rmdefault}{\mddefault}{\updefault}$x$}}}}
\put(3276,-2099){\makebox(0,0)[lb]{\smash{{\SetFigFont{9}{10.8}{\rmdefault}{\mddefault}{\updefault}$f(x+n_{1})$}}}}
\put(5376,-2124){\makebox(0,0)[lb]{\smash{{\SetFigFont{9}{10.8}{\rmdefault}{\mddefault}{\updefault}$y=f(x+n_{1})+n_{2}$}}}}
\end{picture}%
$~~~~~~~~$}},{\centerline{\small{Elementary Relay Channel}}}]
\end{window}
There is no direct link between the source and the destination,
which may be due to the half duplex constraint of the nodes, where
in the first slot D serves a different set of nodes. The transmit
power at the source and the relay is $P$ and $P_{R}$ respectively.
At both the relay and the destination, the received symbol is
corrupted by additive white Gaussian noise of unit power. Relay R
observes $r$, a noisy version of the transmitted symbol $x$. Based
on the observation $r$, the relay transmits a symbol $f(r)$ which is
received at the destination along with its noise $n_{2}$.
\begin{eqnarray}
r&=&x+n_{1}\nonumber\\
y&=&f(r)+n_{2}
\end{eqnarray}
The relay function $f$ satisfies the average power constraint,
\textit{i.e.} $\mathcal{E}_{r}\left[f(r)|^{2}\right]=P_{R}$. Without
loss of generality, channel gains for the source-relay and the
relay-destination link can be incorporated into the model by
modifying $P$ and $P_{R}$ appropriately.
 We seek to determine the memoryless
relay function $f(.)$ that maximizes SNR at the destination.

\subsection{What is the definition of SNR?}
Given an observation $y$, that contains a desired signal $x$ as well
as some distortion (noise), SNR is traditionally defined as the
power of the signal $P_x$ divided by the power in the noise
component $P_n$. For observations of the form $y=x+n$ where the
observed power $P_y = P_x + P_n$ (i.e. signal and noise are
uncorrelated) it is easy to separately identify the contribution of
the signal power and the noise power to the observed power. However,
what is the definition of SNR if the observation $y$ is not already
explicitly presented in the standard form $y=x+n$ with uncorrelated
signal and noise components? In general, the observation $y$ may
have an arbitrary and possibly non-linear dependence on the desired
signal $x$. For example, consider the signal at the destination: $y
= f(x+n_1) + n_2$ with an arbitrary function $f()$ describing the
memoryless relay functionality. In order to define SNR one needs to
separately identify the power contributions of the signal and noise
components to the observed signal $y$. If we can identify $P_y = P_x
+ P_n$ then the definition of SNR readily follows as
$\frac{P_x}{P_n}$. In other words, the definition of SNR follows
from a representation of the observation $y$ in the form $y = x +
n$, with \emph{uncorrelated} signal and noise components. To this
end, we view the signal $y$ as a scaled version of the sum of the
signal $x$ and an error component $e_{u}$ uncorrelated with $x$.
\begin{equation}
y=f(x+n)=\frac{\mathcal{E}[x^*y]}{\mathcal{E}[|x|^2]}(x+e_{u}).
\label{uncorr_error_relation}
\end{equation}
Notice that any signal $y$ can be expressed as above regardless of
whether $y$ is a linear or non-linear function of $x+n$. Rearranging
(\ref{uncorr_error_relation}),
\begin{equation}
e_{u}=\frac{\mathcal{E}[|x|^2]}{\mathcal{E}[x^*y]}y -x.
\label{uncorr_error_relation1}
\end{equation}
It is easy to verify that $e_{u}$ is uncorrelated to $x$.
\begin{eqnarray}
\mathcal{E}[x^{*}e_{u}]&=&\mathcal{E}\left[x^{*}\left(\frac{\mathcal{E}[|x|^2]}{\mathcal{E}[x^*y]}y-x\right)\right]\\
&=&\frac{\mathcal{E}[|x|^2]}{\mathcal{E}[x^*y]}\mathcal{E}[x^{*}y]-\mathcal{E}[x^{*}x]=0
\end{eqnarray}
To calculate the SNR of the received signal $y$, we need to identify
the error term in the received signal $y$, that is uncorrelated to
the signal $x$. The scaling factor
$\frac{\mathcal{E}[x^*y]}{\mathcal{E}[|x|^2]}$ in
(\ref{uncorr_error_relation}) is common to both the signal and error
terms. Therefore, the generalized SNR is defined as follows:
\begin{equation}
\mbox{GSNR}=\frac{\mathcal{E}[|x|^2]}{\mathcal{E}[|e_{u}|^2]}=\frac{\mathcal{E}[|x|^2]}{\mathcal{E}[|\alpha
y-x|^2]}, \label{GSNR_defn}
\end{equation}
where $\alpha=\frac{\mathcal{E}[|x|^2]}{\mathcal{E}[x^*y]}$.  The
advantage of the generalized definition lies is its applicability to
both linear and nonlinear relay functions. Note that the
conventional definition of SNR for point to point links is a special
case of the generalized SNR. For example, consider a received signal
$y=hx+n$. The conventional SNR is $|h|^2P$. To obtain the
generalized SNR, we need to express $y$ as a sum of the signal $x$
and uncorrelated error $e_{u}$ in the following form.
\[y=\frac{\mathcal{E}[x^*y]}{\mathcal{E}[|x|^2]}(x+e_{u}), \]
where $\frac{\mathcal{E}[x^*y]}{\mathcal{E}[|x|^2]}=h$, in this
case. Therefore $e_{u}=\frac{n}{h}$. The generalized SNR from
(\ref{GSNR_defn}) is,
\[\mbox{GSNR}=\frac{P}{\frac{1}{|h|^2}}=|h|^2P,\] which is also the
conventional definition of SNR.

The GSNR concept can be viewed as a decomposition of an observation
into a component along the desired signal space and its orthogonal
signal  (uncorrelated noise) space. The orthogonal projections are
evident in the second moment constraint $P_y=P_x+P_n$ (Pythagoras
Theorem). GSNR is therefore as natural and meaningful a metric as
the orthogonal projections themselves. While GSNR optimization does
not guarantee capacity or BER optimality it is interesting to note
that all three metrics(BER, capacity, GSNR) lead to very similar
optimal relay functions for BPSK. The BER optimal Lambert-W function
is very similar to the GSNR optimal tan-hyperbolic function(EF).
Moreover, in a separate work we have shown that, numerically, the
GSNR optimal EF function is also capacity optimal for BPSK
\cite{cap_icc06}. To summarize, GSNR optimality is related to
capacity and BER optimality and offers a tractable performance
optimization metric.
\subsection{Optimal Relay Function}
We first derive the optimal estimation method that maximizes GSNR.
Based on this result, we determine the optimal relay function.
\begin{theorem}
\label{SNR_MAX_EST_THEOREM} Given an observation $r$ that contains
both the signal $x$ and noise $n$, the MMSUE (SNR maximizing
estimate) of $x$ is
\[\hat{X}(r)=\frac{\mathcal{E}[|x|^2]}{\mathcal{E}[x^*\mathcal{E}(x|r)]}\mathcal{E}[x|r],\]
regardless of the input and the noise distributions.
\end{theorem}
\verb"Proof": Without loss of generality, any estimator $\hat{X}(r)$
can be expressed as
\[\hat{X}(r)=x+e_{u},\] where $e_{u}$ is uncorrelated with $x$.
It is clear from (\ref{GSNR_defn}) that minimizing the mean square
uncorrelated estimation error (MMSUEE), $\mathcal{E}[|e_{u}|^2]$
amounts to maximizing SNR. The optimization problem is therefore to
minimize $\mathcal{E}[|e_{u}|^2]$ with respect to $\hat{X}(r)$
subject to the constraint that $e_{u}$ is uncorrelated with $x$. The
constraint is equivalent to $\mathcal{E}[xe_{u}^*]=0$ as
$\mathcal{E}[x]=0$ for all signal constellations. Employing Lagrange
multipliers\footnote{As the constraint $\mathcal{E}[xe_{u}^*]$ is a
complex quantity, the Lagrange multipliers are $\lambda$ and
$\lambda^{*}$ corresponding to $\mathcal{E}[xe_{u}^*]$ and
$\mathcal{E}[x ^*e_{u}]$ respectively.}, we write the constrained
minimization as the minimization of
\begin{eqnarray*}E&=& \mathcal{E}[|e_{u}|^2]-\lambda
\mathcal{E}[x^*e_{u}]-\lambda^* \mathcal{E}[xe_{u}^*]
\\
&=&\mathcal{E}[|\hat{X}(r)-x|^2]-\lambda
\mathcal{E}[x^*(\hat{X}(r)-x)]-\lambda^*
\mathcal{E}[x(\hat{X}^*(r)-x^*)]\\&=&\mathcal{E}[|\hat{X}(r)|^2]+(-\lambda+1)
\mathcal{E}[x^*\hat{X}(r)]+(-\lambda^*+1) \mathcal{E}[x\hat{X}^*(r)]-P(-\lambda-\lambda^*-1)\\
&=&\mathcal{E}_{r}\left[|\hat{X}(r)|^2-(\lambda-1) \hat{X}(r)
\mathcal{E}[x^*|r]-(\lambda^*-1)
\hat{X}^*(r)\mathcal{E}[x|r]\right]+P(\lambda+\lambda^*+1)\\
&=&\mathcal{E}_{r}\left[|\hat{X}(r)-(\lambda-1)\mathcal{E}[x|r]|^2-|(\lambda-1)\mathcal{E}[x|r]|^2\right]+P(2\mathfrak{Re}(\lambda)+1)
\end{eqnarray*}
From the above equation, it is clear that $E$ is minimized when
\[\hat{X}(r)=(\lambda-1)\mathcal{E}[x|r],\] and the minimum
mean squared uncorrelated estimation error is
\begin{equation}
E^{*}=P(2\mathfrak{Re}(\lambda)+1)-|\lambda-1|^2|\mathcal{E}_{r}\left[|\mathcal{E}[x|r]|^2\right]\end{equation}
From the constraint that
\[\mathcal{E}[x^*e_{u}]=0 ~~\Rightarrow ~~ \mathcal{E}[x^*\hat{X}(r)]=P, \] we
have
\[\lambda-1=\frac{P}{\mathcal{E}[x^*\mathcal{E}(x|r)]}.\]
Therefore
$\hat{X}(r)=\frac{P}{\mathcal{E}[x^*\mathcal{E}(x|r)]}\mathcal{E}[x|r]$
is the SNR maximizing estimate. %Note that
%$\mathcal{E}[x^*\mathcal{E}(X/R=r)]=\mu + P$, therefore the optimal
%estimate (Minimum Mean Square Uncorrelated Estimate(MMSUE)) is given
%by
%\[\mbox{MMSUE}=\hat{X}(r)=\frac{P}{P+\mu}\mathcal{E}[X/R=r], \]
It should be noted that the above result is completely general and
is valid for all input and noise distributions. \hfill \QED

This result implies that any scaled version of MMSE estimator is
GSNR optimal. Thus regardless of the power constraint at the relay,
EF maximizes GSNR at the output of the relay.

\begin{theorem}
\label{snr_max_scale} In a single relay network, maximizing GSNR at
the output of the relay amounts to maximizing GSNR at the
destination.
\end{theorem}
\verb"Proof": Consider any estimate $\hat{X}(r)=x+e_{u}$, with
$\mathcal{E}\left[|e_{u}|^2\right]=E$, the uncorrelated estimation
error power.
\begin{equation}
\mbox{Let}~~~f(r)=\alpha\hat{X}(r)\end{equation} where
$\alpha^2=\frac{P_{R}}{P+E}$ satisfies the relay power constraint.
The received symbol at the destination is
\[y=\alpha\hat{X}(r)+n=\alpha(x+e_{u})+n\] The GSNR at the
destination is given by
\begin{equation}\mbox{GSNR}_{D}=\frac{\alpha^2P}{\alpha^2E+1}=\frac{P}{E+\frac{P+E}{P_{R}}}.\label{snr_des}\end{equation}
Clearly minimizing $E$ amounts to maximizing GSNR at the
destination. \hfill \QED

As scaling does not alter the GSNR of the estimate, forwarding MMSUE
and MMSE results in the same relay function. The fundamental
relationship between these estimation methods is discussed in
Appendix \ref{APP-FUND_RELA}. From the results of theorem
\ref{SNR_MAX_EST_THEOREM} and theorem \ref{snr_max_scale}, we have
the following theorem for the optimal relay function.
\begin{theorem}
For a network with a single relay that has a power constraint
$P_{R}$, the relay function that maximizes GSNR at the destination
is
\[f(r)=\sqrt{\frac{P_{R}}{\mathcal{E}_{r}[\left|\mathcal{E}(x|r)\right|^2]}}\mathcal{E}[x\mathcal{|}r],\]
regardless of the input and noise distributions.
\label{THEOREM_OPT_FN}
\end{theorem}

Thus the new memoryless forwarding strategy, estimate and forward is
GSNR optimal in a single relay network. In the next section, we
compare the performance

\section{Comparative Analysis} \label{comparative section}
From the mean square uncorrelated estimation error at the relay,
the SNR at the destination for any forwarding scheme can be
obtained directly from (\ref{snr_des}). Therefore calculation of
uncorrelated error power of DF and AF allows a direct comparison
of these schemes with the SNR optimal EF. To determine the
estimate whose error is uncorrelated to the signal $x$ from the
actual relay function, we only need to obtain the scaling factor
that allows the relay function to be expressed as in
(\ref{uncorr_error_relation}). The relay function for DF depends
on the modulation scheme as discussed earlier. In this section, we
compare the schemes for BPSK modulation and illustrate the concept
of generalized SNR.
\subsection{Demodulate and Forward}
We express the relay function of a demodulating relay as
\[f_{DF}(x+n)=\sqrt{P_{R}}\mbox{sign}(x+n)=\sqrt{\frac{P_{R}}{P}}(x+d)\]
where $d$ is the Euclidean distance between the input symbol $x$ and
the demodulated symbol. The distribution of $d$ conditioned on $x$
is given by
\begin{equation}
d=\left\{
            \begin{array}{lr}
            0 & 1-\epsilon \\
            -2x & \epsilon
            \end{array}
            \right. \label{df_distr}
\end{equation}
where $\epsilon=Q\left(\sqrt{P}\right)$, the probability of symbol
error. As seen from the error distribution, the demodulation error
$d$ is correlated with $x$. The correlation between the input and
the error is given by
\begin{equation}
\mathcal{E}(xd)=-2P\epsilon. \label{dfcorr}
\end{equation}
The uncorrelated error can be calculated from
(\ref{uncorr_error_relation}) according to which
\begin{equation}e_{u}=\frac{P}{\mathcal{E}(xf_{DF}(x+n))}f_{DF}(x+n)-x=\frac{P}{P-2P\epsilon}(x+d)\label{df_msue}\end{equation}
From (\ref{dfcorr}), the power of the uncorrelated error in
(\ref{df_msue}) can be calculated and is given by
\begin{equation}
\mbox{MSUEE}_{DF}=\frac{4P\epsilon (1-\epsilon)}{(1-2\epsilon)^2}
\end{equation}
To characterize the mean squared uncorrelated error at the output of
the DF relay, we first consider $\epsilon$, the probability of
decision error at the relay.
\[\epsilon=Q\left(\sqrt{P}\right)=\frac{1}{\sqrt{2\pi}}\int_{\sqrt{P}}^{\infty}\exp(-\frac{x^2}{2})dx=\frac{1}{2}-\frac{1}{\sqrt{2\pi}}\int_{0}^{\sqrt{P}}\exp(-\frac{x^2}{2})dx\]
For small values of $P$,
\[\epsilon=\frac{1}{2}-\sqrt{\frac{P}{2\pi}}\]
Therefore at low source transmit power, the mean squared
uncorrelated error can be expressed as a function of $P$.
\begin{equation}
\mbox{MSUEE}_{DF}(P)=\frac{4P\left(\frac{1}{2}-\sqrt{\frac{P}{2\pi}}\right)
\left(\frac{1}{2}+\sqrt{\frac{P}{2\pi}}\right)}{\left(1-\left(1-\sqrt{\frac{2P}{\pi}}\right)\right)^2}=2\pi\left[\frac{1}{4}-\frac{P}{2\pi}\right]
\end{equation}
As $P\rightarrow 0$, the uncorrelated error power shoots up to
$\frac{\pi}{2}$. It should be noted that the noise variance at the
relay is $1$. This suggests that DF is not preferable at low $P$.
\subsection{Amplify and Forward} As the relay function of an AF relay is a scaled
version of the received signal $r$, it is simple to determine the
mean squared uncorrelated error. From (\ref{uncorr_error_relation}),
we have
\[e_{u}=\frac{P}{\mathcal{E}\left[x^*f_{AF}(x+n)\right]}f_{AF}(x+n)-x=n\]
The uncorrelated error power is therefore the same as the noise
variance, $\mbox{MSUEE}_{AF}=1$, interestingly independent of the
source transmit power.
\begin{figure}
\begin{center}
\centerline{\psfig{figure=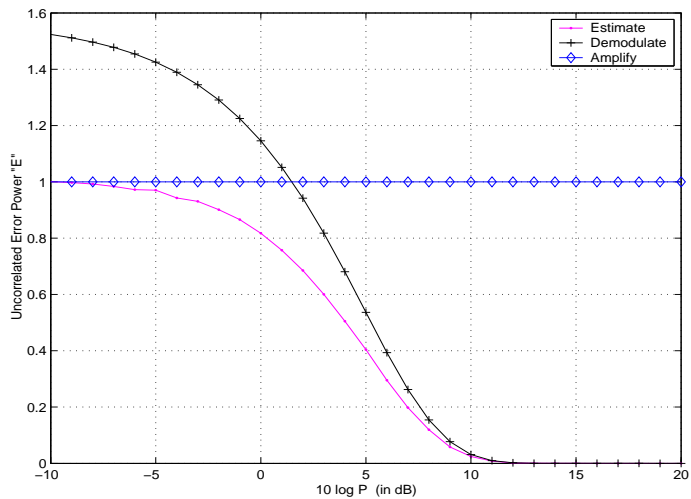,width=3.6in, height=2.6in}}
\caption{Uncorrelated error power vs transmit power with BPSK
modulation} \label{uncorr}
\end{center}
\end{figure}

Fig. \ref{uncorr} plots the uncorrelated estimation error as a
function of transmit power for all the three schemes. Several
interesting observations can be made. It can be seen that AF is
close to optimum (EF) at low $P$ while DF is near optimal at high
$P$. In the intermediate range, both AF and DF are far from optimal.
It is well known that AF suffers from noise amplification at low SNR
\cite{fwd_strategies_yates}, which is in contrast to the results
here. When we view the relay operation as an estimation, it is only
natural that the estimation error is high at low $P$, which results
in noise amplification. In fact AF is very close to optimum among
all memoryless function at low $P$. Rather it is DF that suffers the
most from noise/error\footnote{the term `error' is more appropriate
as noise process is usually independent of the input} amplification.
However AF is inefficient at high $P$ as $\mbox{MSUEE}_{AF}=1$ does
not decrease with $P$, while uncorrelated error in DF and EF
vanishes at high $P$. %For large $P$, the relay functions of EF and
%DF are quite similar. Similarly, the relay function of EF is more
%linear at low SNR
The mean squared uncorrelated error power of the schemes for extreme
values of $P$ is listed in Table \ref{table}.
\begin{table}[hbt]
\begin{center}
\begin{tabular}{|c||c|c|}
\hline \multicolumn{3}{|r|}{~ MMSUE~~~~~~~~}\\ \hline
Relay Function & $P\rightarrow 0$& $P\rightarrow \infty$ \\
\hline  Amplify & $1$& $1$ \\
\hline  Demodulate & $\frac{\pi}{2}$& $0$ \\
\hline  Estimate & $1$& $0$ \\
\hline
\end{tabular}
\caption{Uncorrelated error power at output of relay for BPSK
modulation.}\label{table}
\end{center}
\end{table}

%
%\begin{figure}
%\begin{center}
%\centerline{\psfig{figure=ser1,width=3in, height=3in}}
%\caption{Error Probability of forwarding schemes for BPSK
%modulation} \label{ser1}
%\end{center}
%\end{figure}

\subsection{Higher Order Constellations}
\begin{figure}
\centerline{\psfig{figure=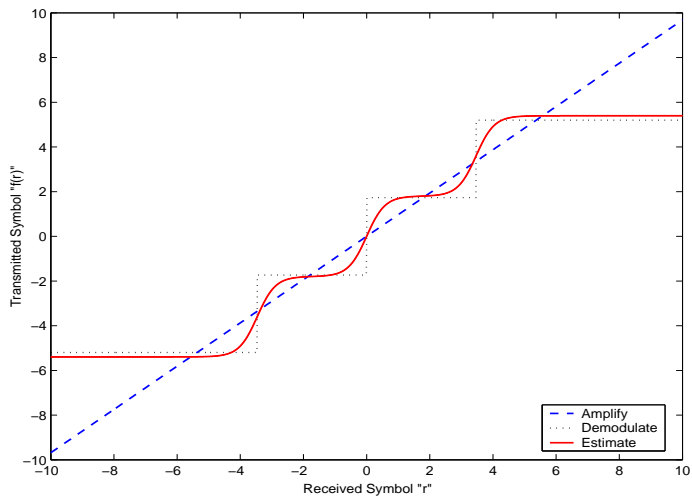,width=3.6in, height=2.6in}}
\caption{Relay Functions for  4-PAM modulation} \label{pam_4}
\end{figure}

We know from theorem \ref{THEOREM_OPT_FN} that EF is GSNR optimal
for all modulation schemes. For fixed input power $P$, increasing
the number of constellation points $M$ will result in an increased
mean squared uncorrelated power for EF. This is rather intuitive
from the fact that increasing the number of constellation points for
fixed power increases the estimation error. Fig. \ref{pam_4} shows
the relay functions for 4-PAM constellation set. Interestingly, the
relay functions of the schemes become more and more similar with
increase in constellation points.

For Gaussian inputs, the unconstrained MMSE estimate and the linear
MMSE estimate are equivalent.
\[\mathcal{E}[x|r]=\frac{P}{P+1}r\] Thus AF and EF strategies are
the same for a Gaussian source. In this context, it can also be
shown that DF and AF are equivalent for Gaussian inputs. The notion
of demodulation of symbols from a Gaussian source is explained
through the following. A Gaussian distribution is quantized into a
number of states with the probability of the $i^{th}$ state given
by,
\[\mbox{Pr}(x_{i})=\frac{1}{\sqrt{2\pi
P}}\int_{(i-1)\Delta x}^{i\Delta x} \exp\left(\frac{-x^2}{2P}\right)
dx .\] Suppose the source transmits symbols $x_{i}$ according to the
probability distribution above, then the MAP detection rule at the
relay is given by
\[
\widehat{X}(r)=\mbox{arg}\max_{x_{i}}~\mbox{Pr}(x|r)\] In the limit
$\Delta x\rightarrow 0 $, $x$ and $r$ become jointly Gaussian. It is
well known that the conditional mean $\mathcal{E}(x|r)$ maximizes
the joint probability. Therefore $\mathcal{E}(x|r)$ which is also
the MMSE estimate is the output of the ML detector. \textit{Thus for
Gaussian inputs AF, EF and DF are identical}.

\section{Parallel Relay Network}
\begin{figure}[hbt]
\begin{center}
\input{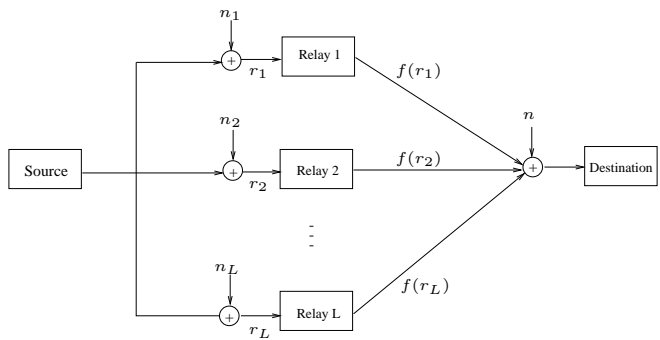}
\caption{Gaussian Parallel Relay Channel} \label{par_nw}
\end{center}
\end{figure}

A Gaussian parallel relay channel \cite{cap_lar_gauss_relay} is
shown in Fig. \ref{par_nw}. It consists of a single source
destination pair with $L$ relays that assist in the communication.
All the links are assumed to be non-fading with unequal channel
gains and information is transferred in two time slots. The relays
observe $\{r_{i}\}_{i=1}^{L}$, the noisy version of the transmitted
signal $x$.
\begin{equation}r_{i}=g_{i}x+n_{i}\end{equation} where
$g_{i}$ is the gain of the link between the source and the $i^{th}$
relay. $n_{i}$ denotes an additive Gaussian noise with $\sigma^2$=1.
Since the relays are assumed to be memoryless, each relay $R_{i}$
transmits a signal that is a function of its observation $r_{i}$. We
assume that the relay function in a parallel relay network is the
same for all the relays, although the channel gains of the relays
may be different. Strictly speaking, an optimal power allocation
based on the channel gains is necessary. However, it is beyond the
scope of the paper. For ease of notation, we denote the transmit
power at the source as $P$ and the relay transmit power as
$\{P_{i}\}_{i=1}^{L}$. Without loss of generality, the channel gain
for the relay destination links can be introduced through the relay
transmit power. The destination receives the sum of all the relay
observations along with its own noise.
\[y=\sum_{i=1}^{L}f(r_{i})+n\]
By viewing relay operation as an estimation we have,
\[f(r_{i})=f(g_{i}x+n_{i})=\alpha_{i}(x+e_{i})\]
where $e_{i}$ is the uncorrelated estimation error at the $i^{th}$
relay, $\alpha_{i}=\sqrt{\frac{P_{i}}{P+E_{i}}}$ and
$E_{i}=\mathcal{E}[\left|e_{i}\right|^2]$, the mean square
uncorrelated error power associated with the relay function. The
received signal at the destination can be expressed as
\begin{equation}
y=\sum_{i=1}^{L}\alpha_{i}(x+e_{i})+n \label{par_rcvd}
\end{equation}

For any forwarding scheme, the SNR at the destination is given by
\begin{equation}
\mbox{GSNR}=\frac{(\sum_{i=1}^{L}\alpha_{i})^2P}{\sum_{i=1}^{L}\alpha_{i}^2E_{i}+\sum_{i=1}^{L}\sum_{j=1,
j\neq i}^{L}\alpha_{i}\alpha_{j}C_{ij}+1},\label{snr_par1}
\end{equation}
where $C_{ij}=\mathcal{E}(e^{*}_{i}e_{j})$ is the correlation
between errors $e_{i}$ and $e_{j}$ at relays $i$ and $j$, $i\neq j$.
For the zero correlation case ($C_{ij}=0,~~ \forall ij$), it is
clear from (\ref{snr_par1}) that the SNR is maximized when the
uncorrelated estimation error at the relays ($E_{i}$) are minimized.
This can also be inferred from the generalized definition of SNR in
(\ref{uncorr_error_relation}). Error in the received symbol at the
destination in (\ref{par_rcvd}) is a linear combination of errors at
the output of relays and the destination noise. When the errors are
uncorrelated, minimizing the error at the output of each of the
relays clearly amounts to maximizing SNR at the destination.

For AF, the correlation $C$ is always zero as the error terms
represent independent AWGN noise. For both DF and EF, the
correlation depends on the modulation scheme and is not always zero.
Although each $E_{i}$ is minimized by EF, due to the possibility of
error correlation, maximum SNR is not always guaranteed. However for
most constellation sets, the error correlation can be shown to be
either very close or exactly equal to zero. Theorem \ref{mpsk_irrot}
in Appendix \ref{rot_subsec} characterizes the rotational property
of estimate and forward for MPSK constellation sets and will be
useful to prove zero correlation property of EF. It proves that, due
to the symmetry in the constellation, MMSE estimate of a signal
rotated by an angle that belongs to a constellation point is the
same as the rotated version of the MMSE estimate of the signal.\\
$~~~~~~~~~~\hat{X}(re^{j\theta_{m}})=e^{j\theta_{m}}\hat{X}(r)$
where $\theta_{m}=\frac{2\pi m}{M}, m=0,1,...M-1$, the signal phases
of MPSK.
\begin{theorem}
\label{corr_mpsk} Error at the relays that estimate and forward are
uncorrelated with each other if MPSK modulation is employed at the
source.\end{theorem} \verb"Proof:" In Appendix
\ref{zero_corr_proof}. Thus $\mathcal{E}[e_{1}e_{2}^*]=0$ for all
MPSK constellation set inputs. This directly suggests that SNR
achieved at the destination is always the highest with estimate and
forward.\hfill \QED

\subsection{Effect of Error Correlation on EF}
The correlation between errors at the output of EF relays, in
general, is not zero for all constellations. However it is
negligible for many constellation sets like M-QAM and it does not
result in any tangible SNR loss. In fact, the correlation can be
expected to decrease for large QAM constellations where the `edge
effects' become insignificant. However due to the combination of
$L(L-1)$ terms for the correlation expression in (\ref{snr_par1}),
we can expect the performance of the system to degrade for very
large values of $L$. In a symmetrical relay network where the
channel gain of all the links are equal, the SNR for any relay
function, as a function of correlation between errors is obtained
from (\ref{snr_par1}) as
\begin{equation}
\mbox{GSNR}=\frac{L^2P}{LE+L(L-1)C+1+\frac{E}{P}},\label{snr_par2}
\end{equation}

\begin{figure}
\centerline{\psfig{figure=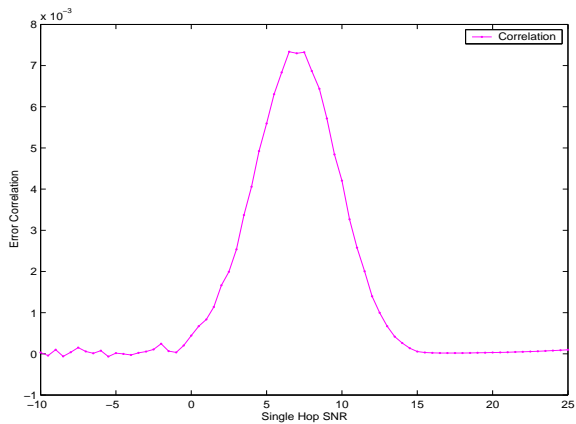,width=3in, height=2.2in}}
\caption{Error Correlation in Estimate and Forward for 16 QAM}
\label{qam_cor}
\end{figure}

Suppose the source employs a modulation scheme that results in
nonzero correlation between errors with EF in a parallel network
with unit channel gains, $\mbox{GSNR}_{AF}>\mbox{GSNR}_{EF}$ only if
\begin{equation}C>\frac{1-E}{L(L-1)}\left(L+\frac{1}{P}\right)\end{equation}
As the scaling associated with the correlation is $L(L-1)$, its
effect is prominent for large $L$. For a given error correlation
$C$, $\mbox{GSNR}_{AF}>\mbox{GSNR}_{EF}$ if
\begin{equation}L \gtrapprox \frac{1-E}{C}+1\end{equation}
The above relation along with Fig. \ref{qam_cor} suggests that, even
if the correlation with EF is nonzero, the number of relays has to
be very large for AF to outperform EF at high SNR, with modulation
schemes like M-QAM.

\subsection{Error Correlation in DF} \label{DF_sec}
Similar to EF, the error correlation in DF depends on the modulation
scheme at the source. For BPSK modulation, the error term in
uncorrelated estimate of DF from (\ref{df_msue}) is
\begin{equation}e_{i}=\frac{d_{i}+2\epsilon_{i}
x}{1-2\epsilon_{i}},\end{equation} where $\epsilon_{i}$ depends on
the transmit power and the source-relay channel. From the
distribution of $d_{i}$ in (\ref{df_distr}), we can calculate the
correlation between errors $e_{i}$ and $e_{j}$ at relay $i$ and $j$.
\begin{eqnarray}C=\mathcal{E}(e_{i}e_{j})&=&
\frac{\mathcal{E}[d_{i}d_{j}]+2\epsilon_{i}\mathcal{E}[xd_{i}]+2\epsilon_{j}\mathcal{E}[xd_{j}]+4\epsilon_{i}\epsilon_{j}\mathcal{E}[|x|^2]}{(1-2\epsilon_{i})(1-2\epsilon_{j})} \nonumber\\
&=&\frac{4\epsilon_{i}\epsilon_{j}P-8\epsilon_{i}\epsilon_{j}P+4\epsilon_{i}\epsilon_{j}P}{(1-2\epsilon_{i})(1-2\epsilon_{j})}=0
\end{eqnarray}
For BPSK modulation and with unit channel gain for all the links,
the effective SNR at the destination is
\begin{equation}
\mbox{GSNR}_{DF}=\frac{PL^2(1-2\epsilon)^2}{4PL\epsilon(1-\epsilon)+1}
\end{equation}

For large M-QAM constellation ignoring edge effects, the
demodulation error can be assumed to be independent of the
transmitted symbol, with the distribution given by
\begin{equation}\small
d=\left\{
            \begin{array}{lr}
            0 & 1-\epsilon \\
            d_{min} & \frac{\epsilon}{4}\\
            -d_{min} & \frac{\epsilon}{4}\\
            jd_{min} & \frac{\epsilon}{4}\\
            -jd_{min} & \frac{\epsilon}{4}\\
            \end{array}
            \right. \label{DF_QAM_DIST}
\end{equation}
where, from \cite{PROAKIS} we have
\begin{eqnarray}\small
d_{min}&=&\sqrt{\frac{6P}{M-1}}\nonumber\\
\epsilon&\leq&4\left(1-\frac{1}{\sqrt{M}}\right)Q\left(\sqrt{\frac{3P}{M-1}}\right)
\end{eqnarray}
Here we assume that decisions error occur only among the nearest
neighbors. Although this is an optimistic assumption, it closely
predicts the performance of the system at medium and high SNR where
the assumption is justified. The effective SNR at the destination
is,
\begin{equation}
\mbox{GSNR}_{DF}\approx\frac{PL^2}{Ld_{min}^2\epsilon+1}.
\label{SNR_Demod}
\end{equation}
\subsection{Asymptotic GSNR Comparison}
While EF is superior to AF and DF at all SNR regardless of the
number of relays, it will be interesting to characterize the
asymptotic gain of EF as a function of L and P. For ease of
analysis, we restrict the channel gains to be equal. From the SNR
expressions of EF and AF, we have the ratio,
\[\frac{\mbox{GSNR}_{EF}}{\mbox{GSNR}_{AF}}=\frac{LP+P+1}{LPE(P)+P+E(P)},\]where
$E(P)$ is the mean squared uncorrelated error power of EF that is a
function of $P$. Note that the above expression does not include the
correlation term. Therefore, it is valid only when there is zero
correlation between the error terms. For any input distribution
$p_{X}(x)$, \textit{i.e.} for all modulation schemes,
\[E(P)\leq 1, ~~\forall P. \]
\subsubsection{Fixed $P$}
With large number of relays ,
\[L \rightarrow \infty,~ \frac{\mbox{GSNR}_{EF}(P)}{\mbox{GSNR}_{AF}(P)}=\frac{MSUEE_{AF}}{MSUEE_{EF}}=\frac{1}{E(P)} ~~~~~~ \frac{\mbox{GSNR}_{EF}(P)}{\mbox{GSNR}_{DF}(P)}=\frac{MSUEE_{DF}}{MSUEE_{EF}}=\frac{4P\epsilon (1-\epsilon)}{(1-2\epsilon)^2E(P)}\]
Notice that MSUEE of the schemes determine the gain. We know from
Section. \ref{comparative section} that $E(P)$ decreases with $P$
and ultimately becomes zero as $P\rightarrow.
 \infty$. This implies that in a large relay network, maximum gain over AF is
obtained for high source transmit power $P$. Similarly maximum
gain over DF is obtained at low $P$. This is due to the fact that
DF is inefficient at low $P$ as indicated by its mean squared
uncorrelated error power.
\subsubsection{Fixed L}
For a fixed number of relays, the GSNR gain of EF over AF at high
$P$ is approximately $L+1$. Similarly the GSNR gain of EF over DF
at low P is very high as indicated in the following expressions.
\begin{equation}P \rightarrow \infty,~~~
\frac{\mbox{GSNR}_{EF}(L)}{\mbox{GSNR}_{AF}(L)}=L+1 ~~~~~~
\frac{\mbox{GSNR}_{EF}(L)}{\mbox{GSNR}_{DF}(L)}=1
\label{asymP1}\end{equation}
\begin{equation}~P \rightarrow 0,~~~~~
\frac{\mbox{GSNR}_{EF}(L)}{\mbox{GSNR}_{AF}(L)}=1~~~
~~~~~~~~~\frac{\mbox{GSNR}_{EF}(L)}{\mbox{GSNR}_{DF}(L)}=\frac{\pi}{2}\label{asymP2}\end{equation}
Above expressions clearly demonstrate the inefficiency of DF and AF
at low and high SNR respectively.
\subsection{Numerical Results }
\begin{figure}
\begin{minipage}[t]{3.3in}
    \centerline{\psfig{figure=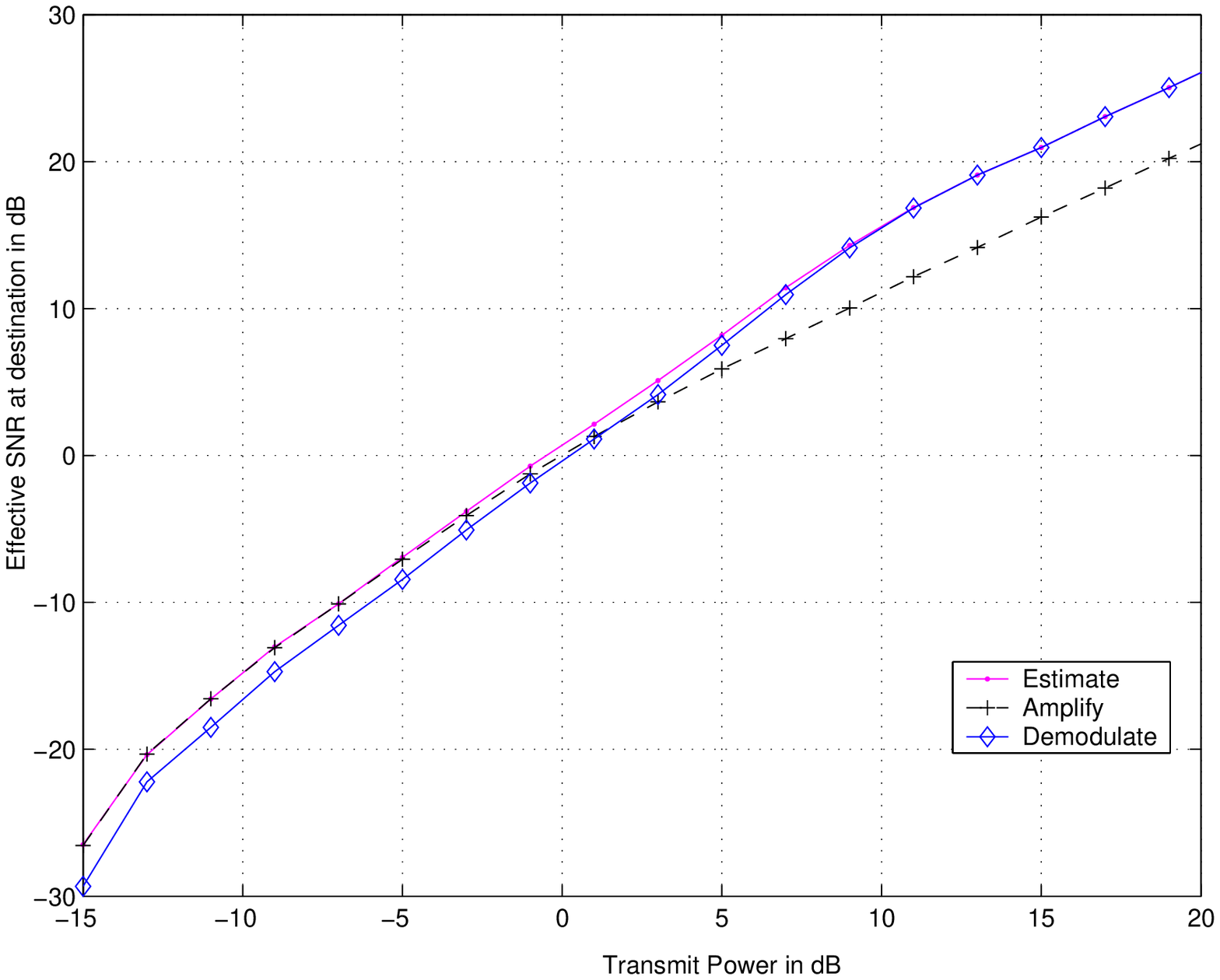,width=3.2in, height=2.7in}}
   \caption{Comparison of SNR at the destination as a function of transmit power $P=P_{R}$ for a parallel network ($L=2$)}
   \label{cap_p2}
\end{minipage}
\hfill
\begin{minipage}[t]{3.3in}
    \centerline{\psfig{figure=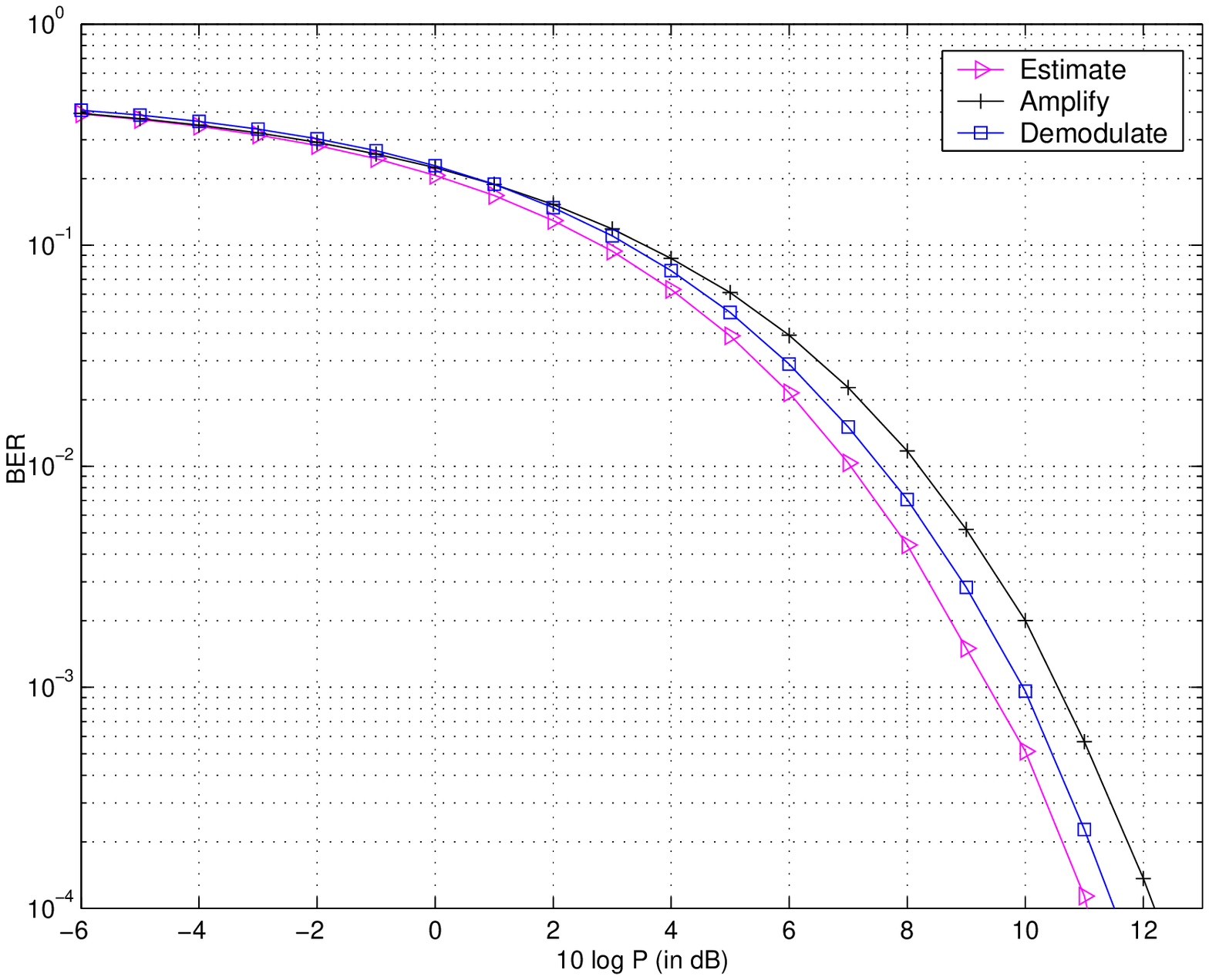,width=3.2in, height=2.7in}}
   \caption{BER of schemes in a parallel network ($L=2$) for BPSK modulation}
   \label{ber_p2}
\end{minipage}
\end{figure}

Fig. \ref{cap_p2} provides the GSNR performance of the schemes in a
parallel relay network with equal channel gains. The corresponding
error probabilities are provided in Fig. \ref{ber_p2}. The error
probabilities closely follow the trend exhibited in GSNR. It can be
seen that EF achieves substantial error probability gains over AF
and DF for all values of $P$. As seen in GSNR plot, AF is superior
to DF at low $P$ while DF performs better than AF at high $P$. This
is also indicated by (\ref{asymP1}) and (\ref{asymP2}). For our
system model, the AF relaying scheme is equivalent to the one
proposed in \cite{distributed_multiuser_mmse_berger_wittenben}. It
will be interesting to compare the performance of the schemes with
optimal power allocation similar to
\cite{distributed_multiuser_mmse_berger_wittenben}.

\section{Serial Relay Network}

\begin{figure}[hbt]
\begin{center}
\input{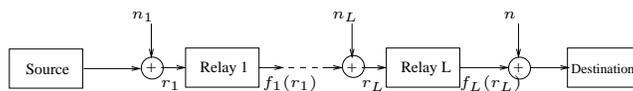}
\caption{Serial Network Model} \label{fig2}
\end{center}
\end{figure}

A serial relay network with Gaussian noise at all receivers in shown
in Fig. \ref{fig2}. All the relays are memoryless and employ a relay
function to transmit a symbol based on its received symbol.  It
should be noted that the relay functions, in general, need not be
the same for all the relays unlike in a parallel network. This is
due to the fact that the noise distribution gets altered at every
hop depending on the relay function of the preceding relay for
multiple relay networks. We assume unit channel gains for the links
and equal transmit power at all the nodes for simplicity of
exposition.
\subsection{Amplify and Forward}
With AF relays in series, the received symbol at the destination can
be expressed as
\begin{equation}
y_{L+1}=\beta^{L}x+\sum_{i=1}^{L}\beta^{i}n_{i} +n.
\end{equation}
\begin{equation}
\mbox{GSNR}_{AF}=\frac{\beta^{2L}P}{1+\sum_{i=1}^{L}\beta^{2i}}
\label{snr_af_ser}
\end{equation}

\subsection{Demodulate and Forward}
 As in Section \ref{DF_sec}, we express the received
signal at the destination as
\begin{equation}
y_{L+1}=x+\sum_{i=1}^{L}d_{i}~+n
\end{equation}
From (\ref{df_distr}) and (\ref{DF_QAM_DIST}), the effective SNR at
the destination with BPSK modulation and for a large QAM modulation
is obtained.
\[
\mbox{GSNR}_{DF}^{^{\mbox{\tiny
BPSK}}}=\frac{P(1-2\epsilon)^2}{4PL\epsilon(1-\epsilon)+1}~~~~~~~~~~~
\mbox{GSNR}_{DF}^{^{\mbox{\tiny
QAM}}}\approx\frac{P}{Ld_{min}^2\epsilon+1}
\]
\subsection{Estimate and Forward}
With estimate and forward at all the relays, the corresponding relay
functions varies with each relay as the noise distribution gets
altered at each link due to nonlinear operations performed at the
preceding relay. The relay function for the $i^{th}$ is given by
\[f_{i}(r_{i})=\alpha_{i}\mathcal{E}[x|r_{i}=f_{i-1}(r_{i-1})+n_{i}]\]

\begin{proposition}
In a serial relay network, the last relay should perform estimate
and forward for maximizing SNR at the destination, \emph{regardless
of the relay function in the preceding relays}.
\end{proposition}
\verb"Proof:" Regardless of the relay functions at the preceding
relays, the received signal at the last relay can be expressed in
the same form as (\ref{uncorr_error_relation}). From theorem
\ref{SNR_MAX_EST_THEOREM}, which is valid for all input and noise
distributions, it is straightforward that EF at the last relay
maximizes the SNR at the destination. \hfill \QED

\subsection{Performance Comparison}
\begin{figure}
\begin{minipage}[t]{3.3in}
    \centerline{\psfig{figure=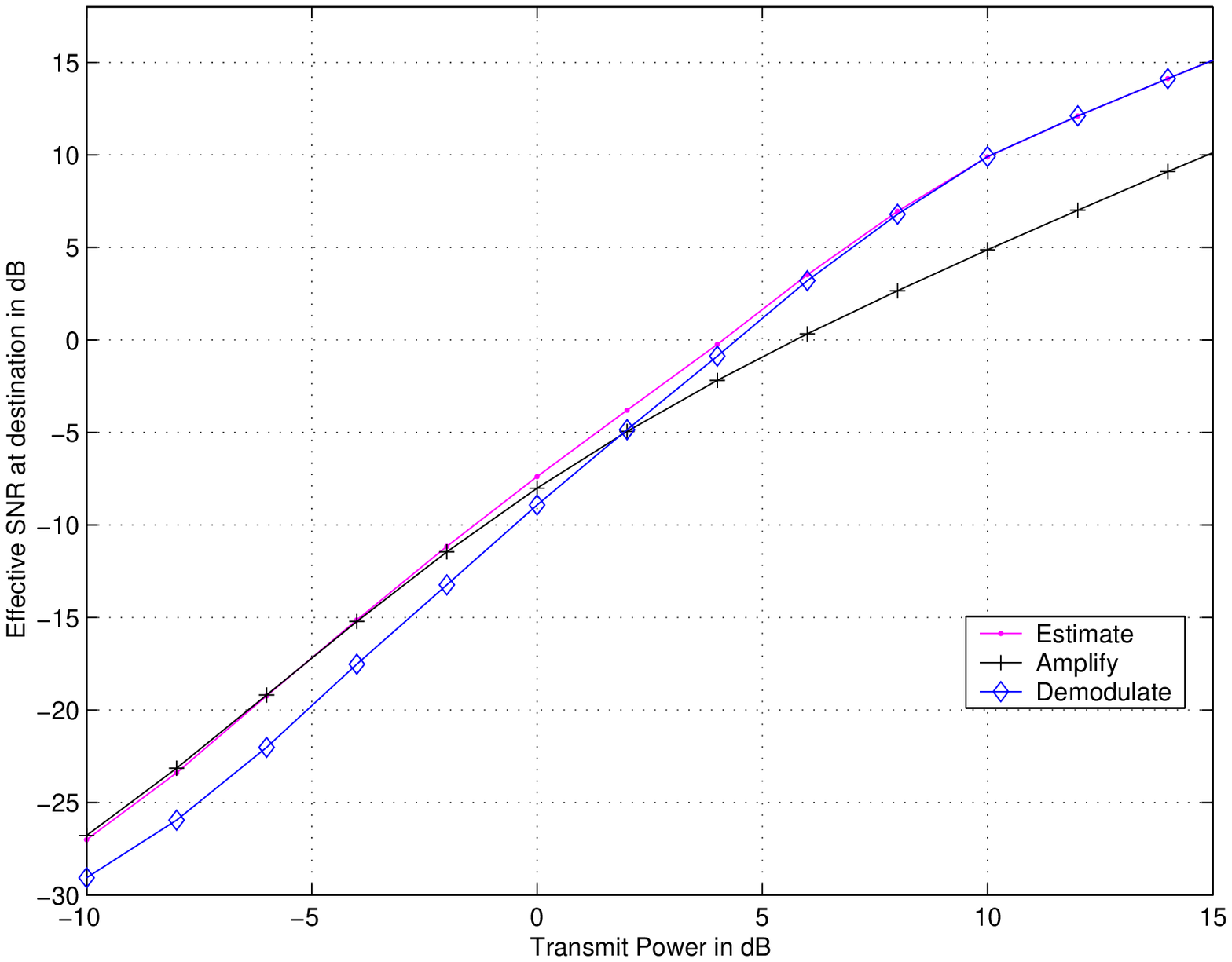,width=3.2in, height=2.7in}}
   \caption{Comparison of SNR at the destination as a function of transmit power $P=P_{R}$ for a serial network ($L=2$)}
   \label{cap_s2}
\end{minipage}
\hfill
\begin{minipage}[t]{3.3in}
    \centerline{\psfig{figure=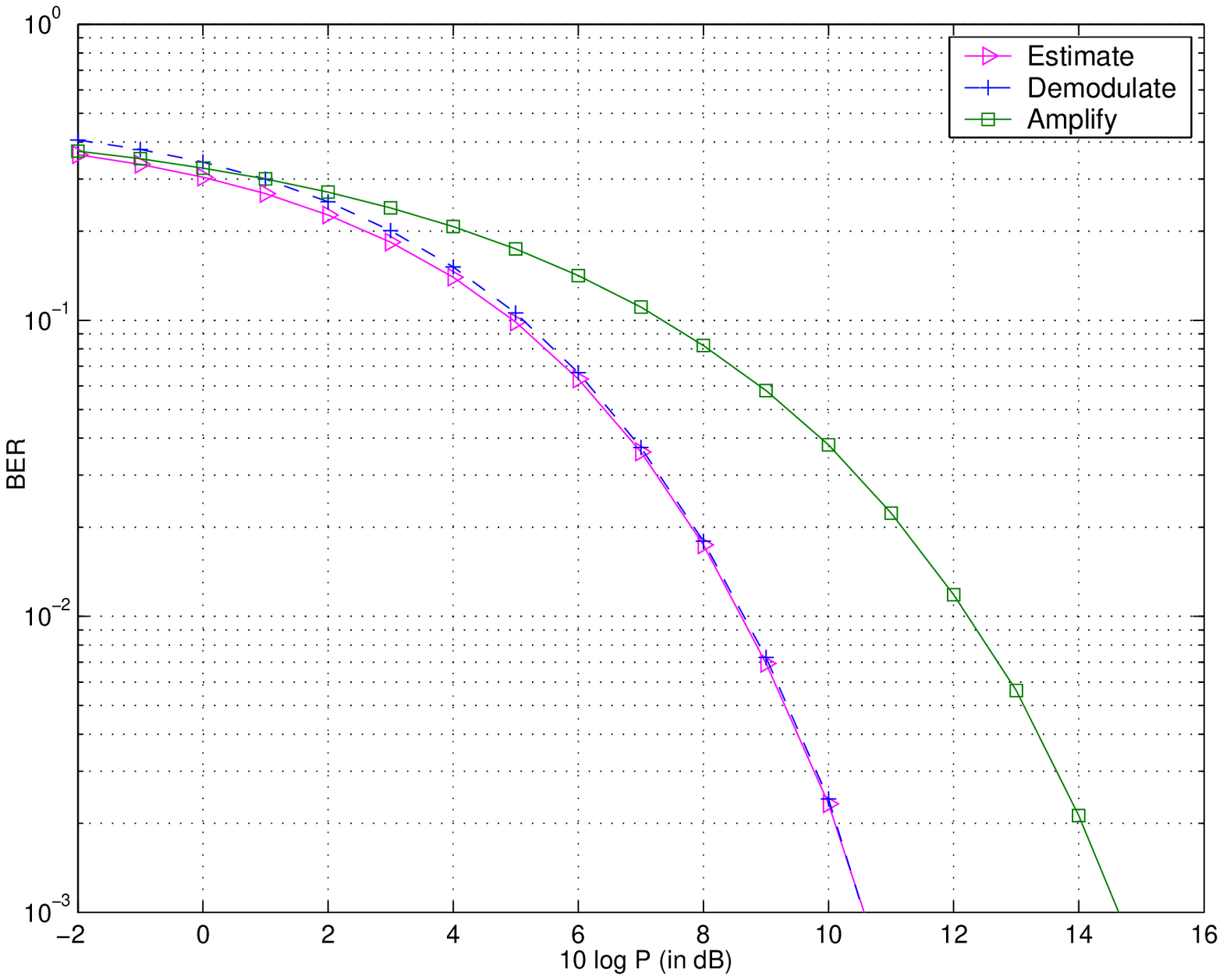,width=3.2in, height=2.7in}}
   \caption{BER of schemes in a parallel network ($L=2$) for BPSK modulation}
   \label{ber_s2}
\end{minipage}
\end{figure}

Fig. \ref{cap_s2} compares the destination SNR of the schemes for
two serial relays. Here the relay functions for DF and AF remain the
same for both the relays. For EF,
$f_{1}(r_{1})=\alpha_{1}\tanh(\sqrt{P}r_{1})$ and
$f_{2}(r_{2})=\alpha_{2}\mathcal{E}[x|r_{2}=\alpha_{1}\tanh(\sqrt{P}r_{1})+n_{2}]$.
As expected, EF is the best performing scheme and DF closely follows
it.

It can be easily noticed that in a serial network, the effective SNR
decreases with each stage. AF, being power inefficient at high SNR,
suffers the most due to multi-hop communication.
\begin{equation}\mbox{GSNR}_{AF}=\frac{P}{1+\sum_{k=1}^{L}\frac{1}{\beta^{2k}}}< \frac{P}{L+1}\end{equation}
For large M-QAM modulation, effective SNR at the destination with DF
scheme can be approximated as
\begin{equation}
\mbox{GSNR}_{DF}=\frac{P}{Ld_{min}^2\epsilon+1}
\end{equation}
Clearly when $d_{min}^2\epsilon<1$ (at high SNR regime),
\begin{equation}
\mbox{GSNR}_{DF}\geq \frac{P}{L+1},
\end{equation}
indicates that DF is superior to AF at high SNR. We can also observe
the case where $\mbox{GSNR}_{AF}>\mbox{GSNR}_{DF}$ at low SNR (when
$d_{min}^2\epsilon> 1$).  Note that the variance of the error
components associated with DF ($d_{min}^2\epsilon$) decreases
exponentially with $P$, while those in AF $(\frac{1}{\beta ^{2k}}$
for the $k^{th}$ relay) decreases linearly with $P$. These
observations can be clearly seen in Fig. \ref{cap_s2}, where AF
performs slightly better than DF at very low SNR. Gradually with
increase in $P$, DF outperforms AF and the performance gap widens
with further increase in $P$.

\subsection{Hybrid Relay Networks}
\begin{figure}[hbt]
\begin{center}
\input{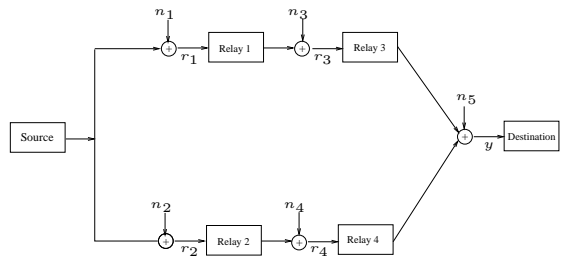}
\caption{Hybrid Serial Parallel Network} \label{ser_par}
\end{center}
\end{figure}

\begin{figure}
\begin{minipage}[t]{3.3in}
   \centerline{\psfig{figure=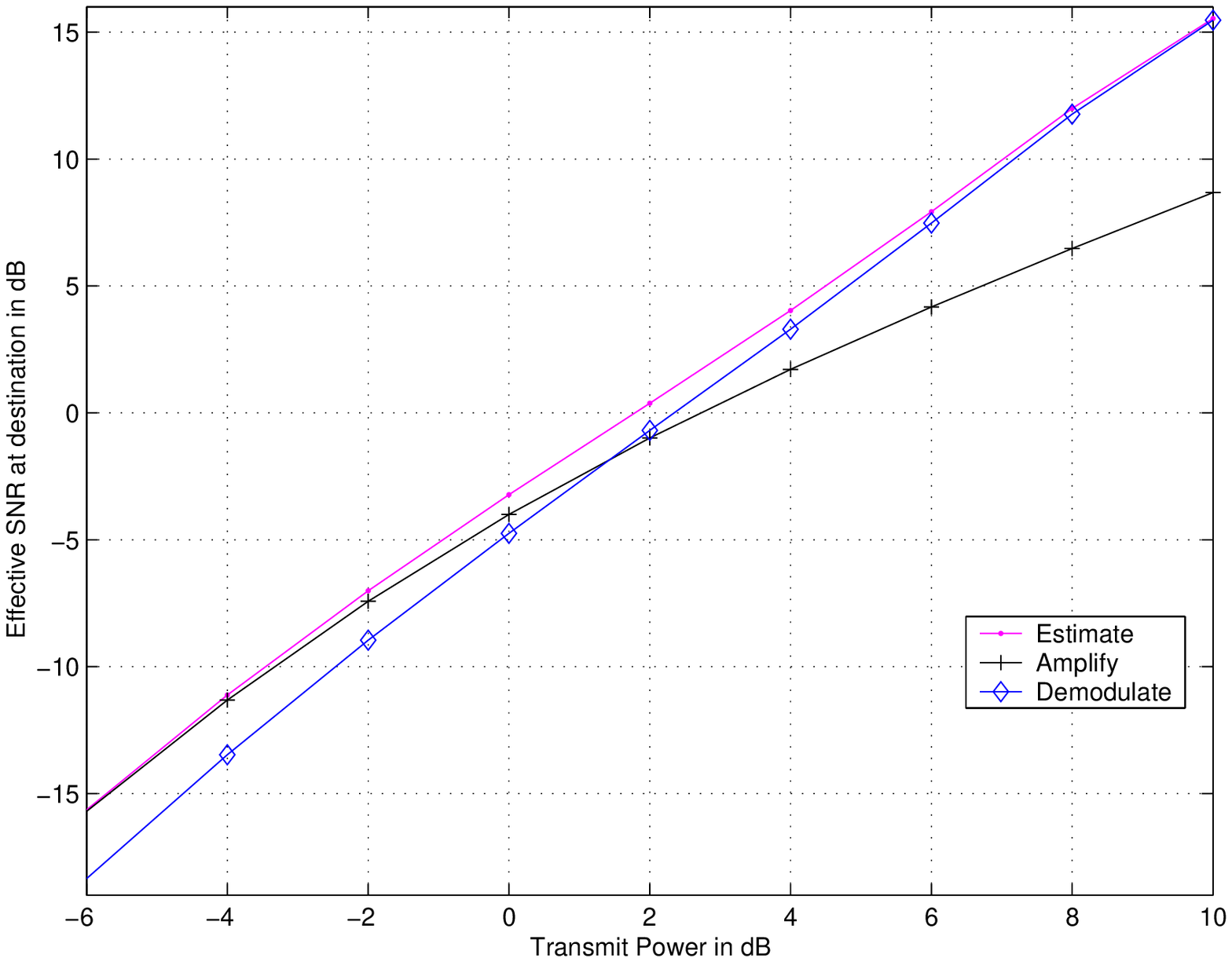,width=3.2in, height=2.7in}}
   \caption{Comparison of SNR at destination as a function of transmit power $P=P_{R}$ for a hybrid network with both serial and parallel elements, for BPSK modulation at source}
   \label{hybrid}
   \end{minipage}
\hfill
\begin{minipage}[t]{3.3in}
   \centerline{\psfig{figure=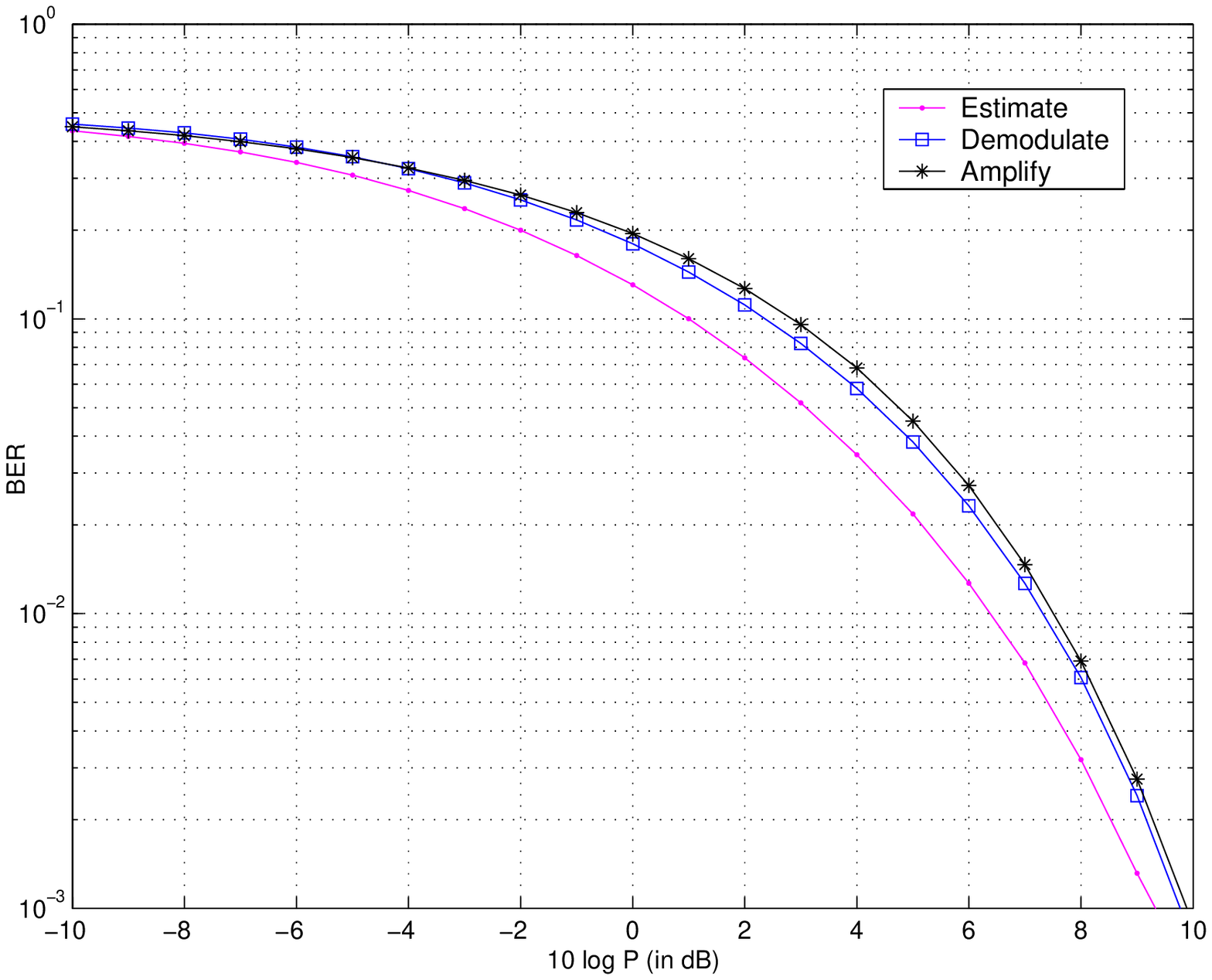,width=3.2in, height=2.7in}}
   \caption{BER of the forwarding schemes for BPSK modulation for the hybrid network in Fig. \ref{ser_par}}
   \label{hybrid_ber}
      \end{minipage}
\end{figure}

From the previous sections, we determine that EF is well suited to
both parallel and serial relay network regardless of $P$, and
substantial performance gain can be obtained over AF and DF in many
scenarios. We also observe that DF is close to optimal in a serial
relay network at high SNR where as AF is near-optimal in parallel
relay networks at low SNR. In these regimes, the performance gain of
EF is limited. Thus, it is interesting to determine the performance
gain of EF in general memoryless relay networks. Consider a network
consisting of both parallel and serial subnetworks as shown in Fig.
\ref{ser_par}. Due to the presence of parallel and serial elements
together in the network, we find a significant performance
degradation in both AF and DF at all $P$. Precisely, this is a
scenario where EF obtains a large gain over the best of DF and AF.
Fig. \ref{hybrid} compares the performance of schemes for the hybrid
network in Fig. \ref{ser_par}. It can be noticed that EF performs
significantly better than the best of DF and AF. Fig.
\ref{hybrid_ber} displays the error probability of the schemes for
the hybrid network. It can be seen that substantial gain is obtained
over the best of DF and AF. The performance gain will increase for a
large network with both parallel and serial elements.

\section{Conclusion}
In this work, we address a fundamental problem in relay networks
that involves determining the set of relay functions in a memoryless
relay network that maximizes performance. From an estimation point
of view, we develop a general framework for determining SNR at the
destination for all memoryless relay processing. For the single
relay case, the generalized SNR was shown to be optimized by minimum
mean square uncorrelated error (MMSUE) estimate which is related to
the traditional MMSE estimate by a constant scaling factor. For both
parallel and serial relay networks, we establish the superiority of
EF over DF and AF. We show that, with MPSK modulation at the source,
maximum SNR is obtained at the destination when the relays estimate
and forward, regardless of the number of relays. Further, we
demonstrate that the last stage of a serial relay network must
employ EF for maximizing SNR at the destination. For hybrid networks
that contain both serial and parallel elements, the advantage of EF
over the best of AF and DF is found to be significant.

\begin{appendix}
\subsection{Relation between MMSUEE and MMSEE}\label{APP-FUND_RELA}
Although the relay functions arising out of MMSUE and MMSE estimates
are identical, they are fundamentally distinct as the objectives
optimized by them are different. MMSUE is the minimum achievable
uncorrelated error power while MMSE is the minimum achievable
distortion. By proving that the correlation of the MMSE error $e$
and the input $x$, we obtain the relationship between MMSE and
MMSUE.
\begin{proposition}
\label{neg_corr_prop} Correlation between the MMSE error $e$ and the
input $x$ is always non positive. \end{proposition}(\verb"Proof:" We
express the MMSE estimate as \[\hat{X}(r)=x+e=x+\frac{\mu}{P}
x+e_{u},\] where $\mu=\mathcal{E}[x^*e]$ and $e_{u}$ is uncorrelated
to $x$, \begin{equation} e_{u}=e-\frac{\mu}{P} x
\label{uncorr_estimate_eqn}\end{equation}
 Consider another estimate which is a scaled
version of the MMSE estimate such that
\[\hat{X}_{new}(r)=\frac{\hat{X}(r)}{1+\frac{\mu}{P}}=x+\frac{e_{u}}{1+\frac{\mu}{P}}.\]
As MMSE estimation is optimal distortion minimizing method, we have
the relation
\begin{eqnarray}
\mathcal{E}[|e|^2]&\leq&\frac{1}{\left(1+\frac{\mu}{P}\right)^2}\mathcal{E}[|e_{u}|^2]\\
                  &\leq&\frac{1}{\left(1+\frac{\mu}{P}\right)^2}\left(\mathcal{E}[|e|^2]-\frac{\mu^2}{P}\right)\\
                  &<&\frac{1}{\left(1+\frac{\mu}{P}\right)^2}\mathcal{E}[|e|^2]
                  \end{eqnarray}
which implies $\mu\leq0$. \hfill \QED\\
For Gaussian inputs, a unique relationship between MMSE estimate and
the correlation exists,  which
 is $\mu=\mathcal{E}(Xe)=-\mbox{MMSEE}=\frac{-P}{P+1}$. A direct consequence of the negative correlation of the error with the signal $x$ leads to the following inequality.
\[\mbox{SNR} \leq \frac{\mbox{P}}{\mbox{MMSEE}}.\]
\begin{proposition}
The minimum mean squared uncorrelated estimation error cannot be
less than MMSEE. The precise relationship between MMSUEE and MMSEE
is
\[\mbox{MMSUEE}=\frac{\mbox{MMSEE}-\frac{\mu ^2}{P}}{(1+\frac{\mu}{P})^2}\]
\end{proposition}
\verb"Proof": We have $\mbox{MMSUEE}\geq \mbox{MMSE}$,  by observing
$e_{u}$ to be the distortion arising out of another estimation
method that cannot achieve a mean squared estimation error less than
MMSEE. The exact relationship between the mean square error of these
methods can be obtained from (\ref{uncorr_estimate_eqn}).

\subsection{Rotational Property of EF}
\label{rot_subsec}
\begin{theorem}\label{mpsk_irrot} For all MPSK constellation inputs, the
MMSE estimate has the property
\[\hat{X}(re^{j\theta_{m}})=e^{j\theta_{m}}\hat{X}(r),\] where
$\theta_{m}=\frac{2\pi m}{M}, m=0,1,...M-1$, the signal phases of
MPSK.
\end{theorem}
\verb"Proof:"
\begin{eqnarray}\mathcal{E}[x|r=re^{j\theta_{m}}]&=&\frac{\sqrt{P}}{M}
\sum_{k=0}^{M-1}
 e^{j\theta_{i}} \mbox{Pr}[x=\sqrt{P}e^{\theta_{k}}|re^{j\theta_{m}}]\\
 &=&\frac{\sqrt{P}}{M}
\sum_{k=0}^{M-1}
 e^{j\theta_{k}} \mbox{Pr}[x=\sqrt{P}e^{\theta_{i}}e^{-j\theta_{m}}|r]\\
 &=&\frac{\sqrt{P}}{M}
 e^{j\theta_{m}}\sum_{k=0}^{M-1}
 e^{j\theta_{(k-m)}}
 \mbox{Pr}[x=\sqrt{P}e^{\theta_{{(k-m)}}}|r]\\
&=& e^{j\theta_{m}}\mathcal{E}[x|r]\nonumber
\end{eqnarray} \hfill\QED
\subsection{Proof for Zero Error Correlation of EF}\label{zero_corr_proof}
Expressing error as the difference of the estimate and the actual
symbol, we have\begin{eqnarray}
C=\mathcal{E}[e_{1}e_{2}^*]&=&\mathcal{E}[(\hat{X}(r_{1})-x)(\hat{X}^{*}(r_{2})-x^{*})]\label{corr_final_eqn}
\\
&=&\left[\frac{P^2}{\mathcal{E}[x^*\mathcal{E}(x|r_1=g_{1}x+n_{1})]\mathcal{E}[x^*\mathcal{E}^{*}(x|r_2=g_{2}x+n_{1})]}\mathcal{E}\left[\mathcal{E}[x|r_{1}]\mathcal{E}[x^{*}|r_{2}]\right]\right]-P
\nonumber
\end{eqnarray}
%&=&\left[\frac{P^2}{\left|\sum_{m=1}^{M-1}\frac{\sqrt{P}}{M}e^{-j\theta_{m}}\mathcal{E}(x|r=r_i)]\right|^2}\mathcal{E}\left[\mathcal{E}[x|r=r_{1}]\mathcal{E}[X^{*}/R=r_{2}]\right]\right]-P\\
%\end{eqnarray}
\begin{eqnarray}
\mathcal{E}\left[\mathcal{E}[x|r_{1}]\mathcal{E}[x^{*}|r_{2}]\right]
&=&\frac{1}{M}\sum_{i=0}^{M-1}\mathcal{E}_{n_{1}}\mathcal{E}[x|r=g_{1}x_{i}+n_{1}]\mathcal{E}_{n_{2}}\mathcal{E}[x^{*}|r_{2}=g_{2}x_{i}+n_{2}]\nonumber\\
&=&\frac{1}{M}\sum_{i=0}^{M-1}\mathcal{E}_{n}\mathcal{E}[x|r=g_{1}x_{i}+n]\mathcal{E}_{n}^{*}\mathcal{E}[x|r=g_{2}x_{i}+n]\label{corr_step_prev}\\
&=&\mathcal{E}_{n}\mathcal{E}[x|r=g_{1}x_{0}+n]\mathcal{E}_{n}^{*}\mathcal{E}[x|r=g_{2}x_{0}+n],
\label{corr_step1}
\end{eqnarray}
where (\ref{corr_step1}) is obtained by applying theorem
\ref{mpsk_irrot} in (\ref{corr_step_prev}).
\begin{eqnarray}
\mathcal{E}[x^*\mathcal{E}(x|r=r_i)&=&\frac{1}{M}\sum_{i=0}^{M-1}x_{i}\mathcal{E}_{n}
\left[\mathcal{E}[x|r=x_{i}+n]\right]\nonumber\\
&=&\sqrt{P}\frac{1}{M}\sum_{i=0}^{M-1}\mathcal{E}_{n}
\left[\mathcal{E}[x|r=x_{0}+n]\right]\label{corr_step2_prev}\\
&=&\sqrt{P}\mathcal{E}_{n}
\left[\mathcal{E}[x|r=g_{1}x_{0}+n]\right] \label{corr_step2},
\end{eqnarray}%
where (\ref{corr_step2_prev}) is reduced to (\ref{corr_step2}) using
theorem \ref{mpsk_irrot}. Substituting (\ref{corr_step1}) and
(\ref{corr_step2}) in (\ref{corr_final_eqn}), we have
\begin{eqnarray}C=\mathcal{E}[e_{1}e_{2}^*]&=&\frac{P^2\mathcal{E}_{n}\mathcal{E}[x|r=g_{1}x_{0}+n]\mathcal{E}_{n}^{*}\mathcal{E}[x|r=g_{2}x_{0}+n]}{P\mathcal{E}_{n}
\left[\mathcal{E}[x|r=g_{1}x_{0}+n]\right]\mathcal{E}_{n}^*
\left[\mathcal{E}[x|r=g_{2}x_{0}+n]\right]} -P\nonumber\\ &=&
0\end{eqnarray}
\end{appendix}
\bibliographystyle{ieeetr}
\bibliography{biblio}
\end{document}